\documentclass[prl,twocolumn]{revtex4}
\usepackage{amsmath}
\usepackage{graphicx,epic,eepic,epsfig,amsmath,latexsym,amssymb,verbatim,subfigure,color}
\usepackage{multirow}
\usepackage{times}
\newcommand{\tr}{{\rm Tr}}

\newcommand{\mE}{{\mathcal E}}

\newcommand{\dk}{x_j}
\newcommand{\idk}{\lfloor x_j\rfloor}
\newcommand{\delt}{\Delta^{n,e}_{t,\sigma}}
\newcommand{\ydelt}{\Delta_{\sigma}^n}
\newcommand{\rf}{\rfloor}\newcommand{\lf}{\lfloor}

\begin{document}
\title{Strengthened quantum Hamming bound}
\begin{abstract}
We report two analytical bounds for quantum error-correcting codes
that do not have preexisting classical counterparts. Firstly the
quantum Hamming and Singleton bounds are combined into a single
tighter bound, and then the combined bound is further strengthened
via the well-known Lloyd's theorem in classical coding theory, which
claims that perfect codes, codes attaining the Hamming bound, do not
exist if the Lloyd's polynomial has some non-integer zeros. Our
bound characterizes quantitatively the improvement over the Hamming
bound via the non-integerness of the zeros of the Lloyd's
polynomial.  In the case of 1-error correcting codes our bound holds
true for impure codes as well, which we conjecture to be always
true, and for stabilizer codes there is a 1-logical-qudit
improvement  for an infinite family of lengths.

\end{abstract}
\author{Sixia Yu$^{1,2}$, C.H. Lai$^{1}$ and C.H. Oh$^{1}$}
\affiliation{$^1$Centre for quantum technologies, National University of Singapore, 3 Science Drive 2, Singapore 117543\\
$^2$Hefei National Laboratory for Physical Sciences at Microscale and Department of Modern Physics, \\
University of Science and Technology of China, Hefei, Anhui 230026, China}

\maketitle
{\it Introduction ---} During all kinds  quantum informational
processes there are uncontrollable noises due to inevitable
interactions with ubiquitous environments. The theory of quantum
error correction \cite{shor1,ste1,ben,knill} provides us a powerful
tool to protect our precious quantum data from various noises by
encoding them into some special subspaces, called the quantum
error-correcting codes (QECCs), which correct certain type of
errors. Stabilizer formalism \cite{gott,cal,kn2,rains0,ket} is one
major method to find such kinds of subspaces, called stabilizer
(additive) codes. There exist also some nonadditive codes
\cite{rains,yu2} with better parameters for whose constructions  a
graphical approach \cite{yu1,loo} as well as an equivalent codeword
stabilizer codes approach \cite{zb} has been developed recently.

One of the fundamental tasks in the theory of QECCs is to find the
optimal codes with as large as possible coding subspaces while
correcting as many as possible errors and consuming as less as
possible resources. Obviously the optimality of the quantum codes,
additive or nonadditive, is subjected to the tradeoffs among those
parameters imposed by the principles of quantum mechanics. The
strongest bound so far is the quantum linear programming (qLP) bound
\cite{qlpbb,rains2} which is unfortunately not analytical. So far
all the analytic bounds for quantum codes have preexisting classical
counterparts \cite{ash}, for examples the quantum versions of
Johnson bound \cite{xian},  Griesmer bound \cite{kp}, and most
importantly, the Singleton bound (qSB) \cite{knill} and the Hamming
bound (qHB) \cite{gott}.

The qHB the qSB are two independent  bounds and comparatively the
qHB is stronger for long codes and weaker for short codes than the
qSB. For single error correcting codes the qHB is quite tight and
can be attained \cite{gott,cal}. For instance in the case of qubits,
i.e., two-level systems, all the optimal stabilizer codes saturating
the qHB are constructed except a few families of lengths
\cite{yu3,jb}. Recently a family of nonadditive codes attaining
asymptotically the qHB has been reported \cite{grsl}. For some
families of lengths however the qHB admits strengthening
\cite{yu3,jb,yu4}, which helps to identify some nonadditive codes
that outperform the optimal stabilizer codes.

Here we will establish two analytical bounds without preexisting
classical counterparts for pure quantum codes.  They are two
strengthenings of the qHB with one arising from an interpolation
with the qSB and the other one from a necessary condition for
perfect codes attaining the Hamming bound. In the case of
1-error-correcting codes we prove that the strengthened qHB holds
true for impure codes as well, which is conjectured to be true for
all distances. Though demonstrated in the case of pure quantum
codes, both bounds apply to corresponding classical codes
straightforwardly.

{\it Quantum Hamming-Singleton bound ---}  In what follows our
physical systems are qudits, $p$-level systems with $p$ being
finite. In the Hilbert space of $n$ qudits  an $K$-dimensional
subspace, whose projection is denoted by $P$, corrects a set of
errors $\{\mE_\omega\}$ if and only if \cite{knill} (in an
equivalent form)
\begin{equation}\label{kl}
P\mE_\omega\mE_\alpha^\dagger P=\frac 1K\tr(P\mE_\omega\mE_\alpha^\dagger)P.
\end{equation}
If a set of errors is correctable then their linear
combinations are also correctable so that we needs only to consider
an operator basis, e.g., qudit Pauli errors $\{\mE_\omega\}$
\cite{kn2}. By a $t$-error  we mean a Pauli error  $\mE_\omega$
acting nontrivially on $\le |\omega|=t$ qudits. As usual
$((n,K,d))_p$ denotes a quantum code of length $n$ (the number of
the physical systems) of size $K$ (the dimension of the coding
subspace) and of distance $d$ meaning that $t$-errors can be
corrected, where $t=\lfloor (d-1)/2\rfloor$.

The quantum  code is said to be {\em pure} or {\em non-degenerate}
if $\tr(P\mE_\omega\mE_\alpha^\dagger)=0$ whenever
$\mE_\omega\not=\mE_\alpha$,  meaning that different errors send the
original coding subspace into different orthogonal subspaces. The
existence of impure codes, where there are errors that we do not
need to worry about, e.g., those errors that stabilize the coding
subspace, makes an essential difference between the classical and
quantum codes.

The quantum  Hamming bound states that if a pure quantum code
$((n,K,d))_p$ exists  then ($d=2t+1+\sigma$ and $\sigma=0,1$)
\begin{equation}\label{hb}
KH_{t,\sigma}^n\le p^n,\quad H_{t,\sigma}^n=p^{2\sigma}\sum_{s=0}^t(p^2-1)^s\binom {n-\sigma} s.
\end{equation}
For odd distances ($\sigma=0$) the qHB can be established by a
standard counting  argument: there are $H_{t,0}^n$ different
$t$-errors and each on of such errors should take the coding
subspace to a different orthogonal subspace. Though the above
counting argument fails for impure codes, there are many positive
evidences \cite{gott,ket,ash,kp,ashly} that the qHB holds true
for impure codes as well.

For even distances $d=2(t+1)$ the  qHB is usually taken as the same
as that for $d=2t+1$.  However a stronger bound as
given in Eq.(\ref{hb}) in the case of $\sigma=1$ can be read off
readily from the classical coding theory, for which there exists
also a simple  counting argument. Consider the set of $(t,1)$-errors
that includes all the $t$-errors and all the $(t+1)$-errors that act nontrivially
on a fixed qudit. It is easy to count that there are
$H_{t,1}^n$ of $(t,1)$-errors and  the product of any two
$(t,1)$-errors is a $(2t+1)$-error. Thanks to the error correction
condition for pure codes  each one $(t,1)$-error should bring the
coding subspace to a different orthogonal subspace so that the qHB
for even distance follows immediately.

In contrast the quantum Singleton bound holds true both for pure and
impure codes and it states that if a quantum code $((n,K,d))_p$
exists then
\begin{equation}\label{sb}
K\le p^{n-2(d-1)}.
\end{equation}
Though several proofs are known, including the original
Knill-Lafflamme argument \cite{knill} and the quantum LP bound
\cite{cal,ket}, we provide here a simple argument which leads in
what follows to a strengthening to qHB. Let us divide $n$ qubits
into 3 subsystems $A,B$, and $C$ containing $d-1$, $d-1$, and
$c=n-2(d-1)$ qudits respectively. By  introducing three reduced
projections $P_A=\tr_{BC} P$, $P_{BC}=\tr_AP$, and $P_B=\tr_{AC}P$
with $P$ being the projection of the coding subspace, we compute
\begin{eqnarray}
\tr P_A^2=\sum_{\alpha\subseteq A}\frac{|\tr(\mE_\alpha P)|^2}{p^{d-1}}=
\sum_{\alpha\subseteq A}\frac{\tr(\mE_\alpha P\mE_\alpha^\dagger P)}{p^{d-1}}\hskip0.5cm
\cr \hskip0.5cm =K\tr P_{BC}^2\ge\frac K {p^c} {\tr P_B^2},
\end{eqnarray}
in which the first equality follows due to the expansion
$P_A\propto\sum_{\alpha\subseteq
A}\tr(P\mE_\alpha)\mE_\alpha^\dagger$ since
$\{\mE_\alpha\}_{\alpha\subseteq A}$ is a basis for subsystem $A$,
the second equality is due to the error-correction condition
Eq.(\ref{kl}), and the inequality is due to the fact $\tr
(P_{BC}-P_B/p^{c})^2\ge 0$. After interchanging subsystems $A$ and
$B$ the above inequality must also hold so that the qSB follows.

The first strengthening  of the qHB comes from an interpolation of
the qSB and qHB for pure codes. Let $0\le e\le t$ be an arbitrary
integer and we divide $n$ physical qudits into two subsystems $A$
and $B$ containing $a=2e$ and $b=n-2e$ qudits respectively. Denote
\begin{equation}
P_e=\sum_{\omega\subseteq B,|\omega|\le t-e+\sigma} \mE_\omega P\mE^\dagger_\omega
\end{equation}
where the summation is taken over all $(t-e)$-errors if $\sigma=0$ while over all the $(t-e,1)$-errors if $\sigma=1$. From the counting arguments for the qHB it follows that $\tr P_e=KH_{t-e,\sigma}^{n-2e}$. On one hand  we have
\begin{equation}
\tr(\tr_AP_e)^2=\frac1{p^{a}}\sum_{\alpha\subseteq A}
\tr(\mE_\alpha P_e\mE^\dagger_\alpha P_e)=\frac{KH_{t-e,\sigma}^{n-2e}}{p^{2e}}
\end{equation}
by noting that $\mE_\omega^\dagger\mE_{\omega^\prime}\mE_\alpha$ is
a  $(d-1)$-error and the error correction condition Eq.(\ref{kl})
for pure codes applies. On the other hand  $\tr(\tr_AP_e)^2\ge
p^{-b}(\tr P_e)^2$ so that we have

{\bf Theorem 1 } If a pure quantum code $((n,K,d))_p$ exists with $d=2t+1+\sigma$ and $\sigma=0,1$ then for all integers $0\le e\le t$
\begin{equation}\label{ib}
KH_{t,\sigma}^{n,e}\le p^n,\quad H_{t,\sigma}^{n,e}=p^{4e}H_{t-e,\sigma}^{n-2e}.
\end{equation}

Some remarks are in order.  i) When $e=t$ or $e=0$ the bound above
coincides with the qSB or the qHB respectively so that our new bound
provides a kind of interpolation and is called here as the {\em
quantum Hamming-Singleton bound} (qHSB).  ii) For a given length $n$
the integer $e$ can be chosen optimally according to
\begin{equation}
e=t+1-\left\lceil\frac{n-d}{p^2-2}\right\rceil,
\end{equation}
with $e=0$ if $n> tq^2+1+\sigma$, i.e., it  coincides with the qHB.
As a result  qHSB improves the qHB for short codes and in fact when
$t(p^2-2)>n-d> p^2-2$ the qHSB is strictly stronger than both the
qSB and qSB. We notice that $n\ge 4(d-1)$ so that there is no
strengthening for qubits, i.e., $p=2$. iii) Recalling that the codes
attaining the qSB and qHB are called as maximum distance separable
(MDS) codes and perfect codes respectively, the qHSB claims that
there is no MDS code \cite{ket} if $n>p^2+d-2$ and no (pure) perfect
code if $n<d+t(p^2-2)$. In the next section the qHSB as well as the
qHB is further strengthened via a necessary condition for the
perfect codes to exist.

{\it Strengthened quantum Hamming bound ---} In the classical coding theory a crucial property of the perfect codes,  codes attaining the Hamming bound, is described by the well-known Lloyd theorem \cite{lloyd} which states that  perfect codes of length $n$ and distance $d=2t+1+\sigma$ with $\sigma=0,1$ do not exist if the Lloyd polynomial
$L^{n}_{t,\sigma}(x)=K_t^{n-\sigma-1}(x-1)$  has some non-integer zeros, where
\begin{equation}
K_t^{n}(x)=\sum_{j=0}^t(p^2-1)^{t-j}(-1)^j\binom {x} j\binom {n-x}{t-j}
\end{equation}
is the Krawtchouk polynomial of degree $t$. In what follows we shall demonstrate a quantitative version of the Lloyd theorem: how the nonintegerness of the zeros of the Lloyd polynomial gives rise to an improvement over the Hamming bound.

Take an arbitrary integer $0\le e< t$ and let $\dk$ be the $j$-th
root ($j=1,2,\ldots,t-e$) of the Lloyd's polynomial
$L_{t-e,\sigma}^{n-2e}(x)$ of degree $t-e$. According to Ref.
\cite{lev} these roots are real and distinct with $0<\dk<n$ with
their integer parts $\lfloor\dk\rfloor$ being different integers. We
introduce the following polynomial
\begin{equation}
\delt(x)=\prod_{j=1}^{t-e}\left(1-\frac x{\idk}\right)\left(1-\frac x{\idk+1}\right)
\end{equation}
of degree $2(t-e)$. It is obvious that $\delt(k)\ge 0$ for integer
$k$ since the roots the polynomial $\delt(x)$ are pairwise
consecutive integers. On the other hand $\delt(\dk)\le 0$ for all $1\le j\le t-e$ and the equality happens for all $j$ only when  all the zeros $\dk$ are integers.

{\bf Theorem 2 } i) If a pure QECC $((n,K,d))_p$ exists with $d=2t+1+\sigma$
where  $\sigma=0,1$ then $S_{t,\sigma}^{n,e} K\le p^n$ for all integers $0\le e< t$ where
\begin{eqnarray}
\frac1{S_{t,\sigma}^{n,e}}=\frac{1}{ H_{t,\sigma}^{n,e}}-\frac{(p^2-1)(n-2e-\sigma)
}{p^{2(2e+1+\sigma)}}\sum_{j=1}^{t-e}\frac{|\delt(\dk)|}{\dk  T_\sigma^t(\dk)}\label{nb}
\end{eqnarray}
with $H_{t,\sigma}^{n,e}$ being the qHSB as defined in
Eq.(\ref{ib}), $\dk$'s being the zeros of Lloyd's polynomial
$L^{n-2e}_{t-e,\sigma}(x)$, and
\begin{equation}
T_\sigma^t(x)=\sum_{s=1}^{t-e}\frac{[K_{s-1}^{n-2e-\sigma-1}(x-1)]^2}{(p^2-1)^{s-1}\binom
{n-2e-\sigma-1}{s-1}}
\end{equation}
being  strictly positive. ii) The above strengthened qHSB is  valid
for impure codes as well for $d=3,4$.

{\bf Proof } We shall postpone the proof of the first part to the
Appendix since it is essentially classical and prove here the second
part which is quantum mechanical since it involves impure codes. In
the case of $d=3,4$, i.e., $t=1$ and $\sigma=0,1$ we have $e=0$ and
for simplicity we denote $S^n_{1\sigma}=S^{n,0}_{1,\sigma}$ and
$\ydelt(x)=\Delta_{1,\sigma}^{n,0}(x)$. The Lloyd polynomial is
linear and its single zero is determined by
$p^2z_\sigma=(p^2-1)(n-\sigma)+1$. For convenience we denote
$\delta_\sigma=z_\sigma-\lfloor z_\sigma\rfloor$,
$\bar\delta_\sigma=1-\delta_\sigma$ and obviously
$\delta_\sigma,\bar\delta_\sigma\ge 0$. The strengthened qHSB is
given by
\begin{equation}
\frac1{S_{1,\sigma}^n}= \frac1{H_{1,\sigma}^n}\left(1-\frac{(p^2-1)(n-\sigma)\bar\delta_\sigma\delta_\sigma}{\lf z_\sigma\rf(\lf z_\sigma\rf+1)}\right).
\end{equation}

According to the  qLP  bound if a code $((n,K,d))_p$ exists then
there is a probability distribution $\{KA_s/p^{n}\}_{s=0}^n$ such
that $\langle K_s^n(x)\rangle=A_s$ for $0\le s\le 2+\sigma$, where
we have denoted $\langle g(x)\rangle=\frac
K{p^n}\sum_{s=0}^ng(s)A_s$ for arbitrary $g(x)$. Since
$\{K_s^n(x)\}_{s\le t}$  provides a basis for the polynomial of
degrees $\le t$ we can expand the polynomial
$\widetilde\Delta_\sigma^n(x)=(n-x)^\sigma\ydelt(x)$ of degree
$2+\sigma$ by $\{K_s^n(x)\}_{s\le 3}$ and, when averaged with
respect to the above probability distribution, we obtain
\begin{equation}
\langle \widetilde\Delta_\sigma^n(x)\rangle=\frac{a_{0\sigma}+a_{1\sigma}A_1+a_{2\sigma}A_2+a_{3\sigma}A_3}{p^{2(1+\sigma)}\lf z_\sigma\rf(\lf z_\sigma\rf+1)}
\end{equation}
in which the coefficients are given by
\begin{eqnarray}
&a_{0\sigma}=n^\sigma(\lf z_\sigma\rf +\bar\delta_\sigma-p^2\delta_\sigma\bar\delta_\sigma),&\cr
&a_{1\sigma}=2(n-1)^\sigma \bar\delta_\sigma+\sigma(\lf z_\sigma\rf+\bar\delta_\sigma-p^2\delta_\sigma\bar\delta_\sigma),&\cr
&a_{2\sigma}=2(n-2)^\sigma/p^2+4\sigma \bar\delta_\sigma,\quad
a_{3\sigma}=6\sigma/p^2.&
\end{eqnarray}
If we are able to
show that $a_{0\sigma}\widetilde\Delta_\sigma^n(i)\ge
a_{i\sigma}\widetilde\Delta_\sigma^n(0)(=n^\sigma a_{i\sigma})$ for all $\sigma=0,1$ and $0\le i\le
2+\sigma$ then the bound $S_{1,\sigma}^nK\le p^n $ follows immediately from
the inequalities
\begin{eqnarray}
\langle  \widetilde\Delta_\sigma^n(x)\rangle&\ge& \frac
K{p^n}\sum_{i=0}^3\widetilde\Delta_\sigma^n(i)A_i \ge \frac {K
n^\sigma}{p^na_{0\sigma}}\sum_{i=0}^{3}a_{i\sigma}A_i\cr &=& \frac
{Kn^\sigma p^{2(1+\sigma)}\lf z_\sigma\rf(\lf
z_\sigma\rf+1)}{p^na_{0\sigma}}
\langle\widetilde\Delta_\sigma^n(x)\rangle.\end{eqnarray}

Suppose  at first $n\ge p^2+2+\sigma$. In this case we have $\lf
z_\sigma\rf\ge p^2$ and, since
$4\delta_\sigma\bar\delta_\sigma\le1$ and
$\delta_\delta,\bar\delta_\sigma\ge0$, $\lf z_\sigma\rf\ge
p^2\delta_\sigma\bar\delta_\sigma+3p^2\bar\delta_\sigma/4$ so that
$a_{0\sigma}\ge 3n^\sigma $ and $a_{0\sigma}\ge 4n^\sigma
\bar\delta_\sigma$. Furthermore it is easy to see that $\ydelt(i)$
with fixed $1\le i\le 2+\sigma$ as a function of $\lf z_\sigma\rf$
is increasing in the range $\lf z_\sigma\rf\ge 4+\sigma$. By
excluding a single case where  $p=2$, $n=7$, and $d=4$ for which the
bound is clear we can assume $\lf z_\sigma\rf \ge 4+\sigma$ so that
$\ydelt(1)\ge (3+\sigma)/(5+\sigma)$, $\ydelt(2)\ge (3+\sigma)/10$,
and $\ydelt(3)\ge 1/5$. If $\sigma=0$ it is now easy to check
$a_{00}\widetilde\Delta_{0}^{n}(i)\ge a_{i0}$ for $i=1,2$. In the
case of $\sigma=1$ it follows from $4n\bar
\delta_1(\widetilde\Delta_1^n(1)-1)\ge 2(n-1)\bar\delta_1=
a_{11}n-a_{01}$ (since $n\ge 7$) that
$a_{01}\widetilde\Delta_1^n(1)\ge na_{11}$. In the mean time from
$a_{01}\ge n(6/p^2+2\bar\delta_1)$,
$6\widetilde\Delta_1^n(2)>2(n-2)$, and
$2\bar\delta_1\widetilde\Delta_1^n(2)\ge 4\bar\delta_1$ as long as
$n\ge 7$ it follows that $a_{01}\widetilde\Delta_1^n(2)> na_{21}$.
Finally from
 $a_{01}\ge 3n$ and $n-3\ge 4$ it follows that
$a_{01}\widetilde\Delta_1^n(3)\ge 12 n/5> 6n/p^2=na_{31}$.

Suppose  now $4+2\sigma\le n\le q^2+1+\sigma$. In this case we have
$\lf z_\sigma\rf=p^2\bar\delta_\sigma=n-\sigma-1$. As a result the
coefficients can be evaluated as
$p^2a_{0\sigma}=n^\sigma(n-\sigma)\lf z_\sigma\rf$,
$p^2a_{1\sigma}=(2+\sigma)(n-1)(n-2)^\sigma$, and
$p^2a_{2\sigma}=(2+4\sigma)(n-2)^\sigma$. It is easy to check
$a_{0\sigma}\widetilde\Delta_\sigma^n(i)\ge n^\sigma a_{i\sigma}$
for $i=1,2,3$. In this case since $p^2a_{0\sigma}=n^\sigma\lf
z_\sigma\rf(\lf z_\sigma\rf+1)$ so that
$S_{1,\sigma}^n=p^{2(2+\sigma)}$ meaning that our strengthened bound
coincides with the qSB $K\le q^{n-2(d-1)}$ in the case of
$d-1=2+\sigma$. \hfill Q.E.D.

Some remarks are now in order. i) To obtain the corresponding
strengthened Hamming bound for classical codes we have only to
replace $K$ by $K/p^{n}$ and then identify $p^2$ with the number of
the alphabet. ii) If the Lloyd's polynomial has at least one noninteger zero
then $|\delt(\dk)|>0$ so that $S_{t,\sigma}^n:=S_{t,\sigma}^{n,0}>H_{t,\sigma}^n$ which yields the Lloyd theorem immediately. iii) When $t=2$
and $n=p^4m((p^2-1)m+1)+2+\sigma$
with $m\ge 0$ all the zeros of the Lloyd's polynomial are
integers and there is no improvement. In the case of $t\ge 3$ (and since $p^2\ge 4$) \cite{hong} there
always exist non-integer zeros for Lloyd's polynomial  so
that we have always $S_{t,\sigma}^{n}>H_{t,\sigma}^n$. iv) Numerical
evidences show that the optimal value of the integer $e$ can be
chosen as follows. Let $\{x_j\}_{j=1}^t$ be the zeros (in an
increasing order) of $L^n_{t,\sigma}(x)$ and let $e=t-j$ with $j$
being the greatest integer such that $\idk< n-d+2j$.

Let us now consider specially the stabilizer code whose size is $K=p^k$ for some integer $k$, which is the number of the logical qudits. The qHB and the strengthened qHB $(e=0)$ for pure stabilizer codes then read $n-k\ge s_p(n,d)\ge h_p(n,d)$ where
\begin{equation}
 h_p(n,d):=\lceil \log_p H_{t,\sigma}^n\rceil, \quad s_p(n,d):=\lceil \log_p S_{t,\sigma}^{n}\rceil.
\end{equation}
As long as $S_{t,\sigma}^{n}>p^{h_p(n,d)}$ we have $s_p(n,d)\ge
h_p(n,d)+1$ since $n,k,h_p(n,d)$ are integers. In this case we say
that the strengthened qHB has a 1-logical-qudit (1-l.q.) improvement
over the qHB. Since $p^2S^n_{t,0}=S^{n+1}_{t,1}$ and
$p^2H^n_{t,0}=H^{n+1}_{t,1}$  a 1-l.q. improvement for the codes of
length $n$ and distance $2t+1$ is equivalent to a 1-l.q. improvement
for the codes of length $n+1$ and distance $2t+2$. In general for a
given $d$ and $n$ satisfying $h_p(n+1,d)=h_p(n,d)+1$ we may expect a
1-l.q. improvement unless all the zeros of the Lloyd
polynomial are integers.

{
 \begin{table}
$\begin{array}{rl@{\hskip 0.1cm}l@{\hskip 0.1cm}l@{\hskip 0.1cm}l@{\hskip 0.1cm}l@{\hskip 0.1cm}l@{\hskip 0.1cm}l@{\hskip 0.1cm}l@{\hskip 0.1cm}l@{\hskip 0.1cm}lllllllllllllccccccccccccc}
\hline
\hline
d&\multicolumn{9}{c}{n_{s_2(n,d)}}\\ [0.5ex]\hline
5&21_{12}&30_{13}&42_{14}&60_{15}&85_{16}&&&&120_{17}\cr
7&25_{17}&31_{18}&39_{19}&49_{20}&61_{21}&62_{21}&78_{22}&98_{23}&123_{24}\cr
9&34_{23}&40_{24}&48_{25}&57_{26}&67_{27}&80_{28}&95_{29}&&113_{30}\cr
11&43_{29}&50_{30}&57_{31}&65_{32}&75_{33}&85_{34}&98_{35}&&112_{36}\cr
13&47_{34}&52_{35}&59_{36}&66_{37}&73_{38}&82_{39}&92_{40}&&103_{41}\cr
15&61_{41}&67_{42}&82_{44}&90_{45}&99_{46}&&&&120_{48}\cr
17&70_{47}&83_{49}&90_{50}&98_{51}&&&107_{52}&116_{53}&127_{54}\cr
19&79_{53}&85_{54}&99_{56}&&&&106_{57}&115_{58}&124_{59}\cr
21&88_{59}&94_{60}&&&&100_{61}&107_{62}&115_{63}&123_{64}\cr
23&&&&&&103_{66}&109_{67}&116_{68}&123_{69}\cr
25&&&&&&&&118_{73}&124_{74}\cr
\hline
\hline
\end{array}$
\caption{Strengthened qHB $s_2(n,d)$ for binary pure stabilizer codes of length
 $n\le 128$ that has a 1-logical-qudit improvement over the qHBs
and meanwhile coincides with the qLP bound.}
\end{table}
}

{\bf Corollary } $s_p(N_{m,\sigma}^r,3+\sigma)= 2(m+1+\sigma)$ while the qHB for the same stabilizer code is $2(m+\sigma)+1$ where
\begin{equation}
N_{m,\sigma}^r=\frac{p^{2m+1}-p}{p^2-1}-r+\sigma \quad (m\ge2)
\end{equation}
with $r$ being an integer satisfying
\begin{equation}
0\le r\le \frac12\left(\sqrt{1-4p^3+4p^4}-p^2-(p-1)^2\right).
\end{equation}

In the case of  $d=4$ and $p=2$ the strengthened qHBs for lengths $N_{m,1}^0$
$(m=2,3,\ldots)$  coincide with the qLP bounds
as far as linear programming can be carried out. Notably there is an
infinite family binary stabilizer codes of lengths $n_a=(4^a-1)/3$
with $a\ge 3$ for which we have  $s_2(n_a,5)> h_2(n_a,5)$. And as far as numerical
calculation is possible, these bounds also coincide with the qLP bounds. For larger distances the numerical results show that the
strengthened qHB for pure stabilizer codes also has a 1-l.q.
improvement over the qHB as tabulated in Table I. There we have
recorded all the strengthened qHBs for the codes of lengths $\le
128$ and odd distances that have a 1-l.q. improvement over the qHBs
and in the same time coincide with the qLP bounds. It seems that for
relative long codes the strengthened qHBs with a 1-l.q. improvement
will coincide with the qLP bounds.

{\it Conclusion and discussion--- } We have established two analytic
bounds for the quantum error-correcting codes that do not have
preexisting classical counterparts. Both bounds strengthen the
quantum Hamming bound for pure codes and in the case of
1-error-correcting codes they are valid also for impure codes as
well. For stabilizer codes of a family of lengths the strengthened
qHB has  a one-logical-qudit improvement over the qHB. Nonadditive
codes that outperform the stabilizer codes for these lengths may be
expected and it is also interesting to find codes that attains the qHSB. Numerical evidences show that for relative large lengths
the strengthened qHB for stabilizer codes with a one-logical-qudit
improvement coincides with the qLP bound no matter how large the
distance is. Furthermore the asymptotic behavior of the strengthened
qHB deserves investigating. Finally we conclude with a conjecture:

{\bf Conjecture }  The strengthened qHSB $KS_{t,\sigma}^{n,e}\le
p^n$ with integer $0\le e\le t$
holds true for impure codes as well for arbitrary 

{\it Acknowledgement ---} The financial support from CQT project
WBS: R-710-000-008-271 is grateful acknowledged.

\newpage

{\it Appendix: proof of Theorem 2 part i) ---}
According to the quantum LP bound if a pure code $((n,K,d))_p$ exists then there exists a probability distribution $\{KA_s/p^n\}_{s=0}^n$ such that
$A_0=1$ and $
\langle K_s^n(x)\rangle=0$ for $1\le s<d$
where we have denoted $\langle g(x)\rangle=\frac K{p^n}\sum_{s=0}^ng(s)A_s$
for an arbitrary $g(x)$.

Any polynomial $g(x)$ of degree $d-1$ has an expansion $g(x)=\sum_{i=0}^{d-1}g_iK_i^n(x)$ with the first coefficient given by
\begin{equation}
g_0=\langle g(x)\rangle_\rho:=\frac 1{p^{2n}}\sum_{s=0}^ng(s)(p^2-1)^s\textstyle\binom ns.
\end{equation}
Due to the qLP bound we have $\langle K_s^n(x)\rangle=0$ for $1\le
s<d$ from which it follows that $\langle g(x)\rangle=\langle
g(x)\rangle_\rho$ as long as $g(x)$ is a polynomial of degree $d-1$.

Take an arbitrary integer $0\le e<t$ and denote $r=2e+\sigma$ for simplicity. We observe that $\binom{n-x}{r} \delt(x)$ is a polynomial of degree $d-1=2t+\sigma$ and thus we have
\begin{eqnarray}\label{lb}
\left\langle\textstyle\binom{n-x}{r} \delt(x)\right\rangle_\rho=\left\langle\textstyle \binom{n-x}{r} \delt(x)\right\rangle
\ge \frac{K}{p^n}\textstyle\binom{n}{r},
\end{eqnarray}
where the inequality is due to $\delt(x)\ge0$ for integer $x$. In order to evaluate the $\rho-$average on the left hand side of  above equation we shall recall some properties of the Krawtchouk polynomials.

First of all we have the Christoffel-Darboux formula \cite{lev}
\begin{eqnarray}\label{cd}
{K^{m}_{t}(y)K^{m}_{t-1}(x)-K^{m}_{t}(x)K^{m}_{t-1}(y)}=\hskip 1.5cm\cr
{p^2(p^2-1)^t\binom m t}\frac{x-y}{t+1}
\sum_{s=0}^{t-1}\frac{K_s^{m}(x)K_s^{m}(y)}{(p^2-1)^s\binom {n} s}.
\end{eqnarray}
Let $\dk$ be the zeros of the Lloyd' polynomial
$L_{t-e,\sigma}^{n-2e}(x)$ and $x_i\not=x_j$ if  $i\ne j$. As a
direct result of Eq.(\ref{cd})  we have
\begin{equation}
\sum_{s=1}^{t-e}\frac{L_{s-1,\sigma}^{n-2e}(x_i)L_{s-1,\sigma}^{n-2e}(x_j)}{(p^2-1)^{s-1}\binom {n-r-1}{s-1}}
=0\end{equation}
Consequently the following polynomial of degree $2(t-e)$
\begin{widetext}
\begin{eqnarray}\label{f}
f(x)&=&{\delt(x)}-\frac{L_{t-e,\sigma}^{n-2e}(x)}{H_{t-e,0}^{n-r}}-
\sum_{j=1}^{t-e}\frac {x\delt(\dk)}{\dk
T_\sigma^t(\dk)}\sum_{s=1}^{t-e}\frac{L_{s-1,\sigma}^{n-2e}(x)L_{s-1,\sigma}^{n-2e}(\dk)}{(p^2-1)^{s-1}\binom
{n-2e-\sigma-1}{s-1}}
\end{eqnarray}
has $x=0$ and $x=\dk$ with $j=1,2,\ldots, t-e$ as zeros so that we can
write $f(x)=x\varphi(x)L^{n-2e}_{t-e,\sigma}(x)$ with
$\varphi(x)$ being some polynomial of degree $t-e-1$. As it turns out, by denoting $\rho_i^n=(p^2-1)^i\binom n i$ for convenience,
\begin{eqnarray}\label{calc}
\left\langle \textstyle\binom{n-x}{r}
\delt(x)\right\rangle_\rho&=&\frac{1}{H_{t-e,0}^{n-r}}{\left\langle
\textstyle\binom{n-x}{r}
L^{n-2e}_{t-e,\sigma}(x)\right\rangle_\rho}+
\sum_{j,s=1}^{t-e}\frac{\delt(\dk) L_{s-1,\sigma}^{n-2e}(\dk)}{\dk
T_\sigma^t(\dk)\rho_{s-1}^{n-r-1}}\left\langle
x\textstyle\binom{n-x}{r}
L_{s-1,\sigma}^{n-2e}(x)\right\rangle_\rho\cr &=&\frac
{\binom{n}{r}}{p^{2r}H_{t-e,0}^{n-r}}+\frac{(p^2-1)(n-r){\textstyle\binom{n}{r}}}{p^{2(r+1)}}\sum_{j=1}^{t-e}\frac{\delt(\dk)
}{\dk T_\sigma^t(\dk)}.
\end{eqnarray}
Because  of $\delt(\dk)\le 0$ for all $j=1,2,\ldots,t$ and
inequality Eq.(\ref{lb}) the strengthened qHSB Eq.(\ref{nb})
follows immediately. To carry out the calculations in
Eq.(\ref{calc}) we have used: i) The following two identities
\begin{eqnarray}
\left\langle\textstyle \binom{n-x}{r}\frac{p^2x}{(p^2-1)(n-r)} \frac{K_{s-1}^{n-r-1}(x-1)}{\rho_{s-1}^{n-r-1}}\right\rangle_\rho=\left\langle\textstyle \binom{n-x}{r}\left(\frac{K_{s-1}^{n-r}(x)}{\rho_{s-1}^{n-r}}-\frac{ K^{n-r}_s(x)}{\rho_s^{n-r}}\right)\right\rangle_\rho
=\frac{\textstyle\binom{n}{r}}{p^{2r}}\delta_{s1}
\end{eqnarray}
\end{widetext}
for $s\ge 1$ and
\begin{eqnarray}
\left\langle\textstyle\binom{n-x}{r} L_{t-e,\sigma}^{n-2e}(x)\right\rangle_\rho={\displaystyle\sum_{s=0}^{t-e}}\left\langle\textstyle\binom{n-x}{r} K_{s}^{n-r}(x)\right\rangle_\rho=\frac{\textstyle\binom{n}{r}}{p^{2r}},
\end{eqnarray}
which follow from  two recurrence relations
\begin{eqnarray}\label{sum}
\label{rc1}
&\frac{p^{2}x}{(p^2-1)n}\frac{K^{n-1}_t(x-1)}{\rho_t^{n-1}}=
\frac{K^n_t(x)}{\rho_t^n}-\frac{K^n_{t+1}(x)}{\rho_{t+1}^n},&\\
\label{rc2}
&{p^{2r}}\textstyle\binom{n-x}{r}K_{s}^{n-r}(x)=\displaystyle\sum_{i=0}^{r}{\textstyle\binom{s+i}i\binom{n-s-i}{r-i}}K_{s+i}^n(x),&
\end{eqnarray}
with the second recurrence relation obtained by computing the coefficients of $u^sv^r$ in $(1-u)^x(1+v+(p^2-1)u)^{n-x}$ in two different ways,
and an identity
\begin{equation}
K_t^{n-1}(x-1)=\displaystyle\sum_{s=0}^tK_s^n(x).
\end{equation}
ii) Due to the recurrence relations Eq.(\ref{rc1}) and
Eq.(\ref{rc2})  the polynomial
$x\binom{n-x}{r}L_{t-e,\sigma}^{n-2e}(x)$ of degree $d$ can be
expanded by $\{K_s^n(x)\mid s\ge t-e\}$. Since the polynomial
$\varphi(x)$ of degree $\le t-e-1$ can be expanded by
$\{K_s^n(x)\mid s\le t-e-1\}$, the orthogonal relations $\langle
K_i^n(x)K_j^n(x)\rangle_\rho\propto\delta_{ij}$ lead to $\langle
\binom{n-x}{r}f(x)\rangle_\rho=0$.

\end{document}